\documentclass[12pt]{iopart}
\usepackage{epsfig}  
\begin{document}

\title[The mystery of the cosmic vacuum energy density]{The mystery of the cosmic vacuum energy density and the accelerated expansion of the Universe \footnote{Invited paper for the European Journal of Physics}}

\author{N. Straumann}

\address {Institute of Theoretical Physics, University of Z\"urich 
\\
Winterthurerstrasse, 190
\\ 8057 Z\"urich (Switzerland)}

\begin{abstract}
After a short history of the $\Lambda$-term it is explained why the 
(effective) cosmological constant is expected to obtain contributions from 
short-distance-physics, corresponding to an energy scale of at least 100 GeV. The actual tiny value of the cosmological constant in any natural scale of units represents, therefore, one of the deepest mysteries of present day fundamental 
physics. We also briefly discuss recent astronomical evidence for a cosmologically significant vacuum energy density causing an accelerating expansion of the universe. This arises mainly from the Hubble diagram of type Ia supernovae and from the observed temperature fluctuations of the cosmic microwave background radiation. If this should become an established fact, we are also confronted with a disturbing {\it cosmic coincidence} problem.
\end{abstract}



\maketitle

\section{Introduction}
Physicists have recognized since quite some time that the smallness of the (effective) cosmological constant is a profound mystery of fundamental physics. Nowadays, the cosmological constant is interpreted as a vacuum energy density. We expect that quantum fluctuations in the fields of the standard model of particle physics, cut off at particle energies of about 100 GeV, contribute to the vacuum energy density. The reason is that there is no symmetry principle in the energy range below the Fermi scale which would require a cancellation of the various contributions.

Based on these considerations we would expect a value for the vacuum energy density which is at least 40 orders of magnitude too large. In spite of some interesting attempts, no convincing solution of this conflict has been offered. Presumably this will only become possible with major advances in fundamental physics. 

There is a second aspect of the cosmological constant problem. Recent measurements of the luminosity-redshift relation for type Ia supernovae strongly suggest a cosmologically significant positive vacuum energy density (or some effective equivalent). The evidence for this is enforced when the supernovae results are combined with measurements of the cosmic microwave background radiation. If this is confirmed we are confronted with the following {\it cosmic coincidence} problem: Since the vacuum energy density is constant in time, while the matter energy density decreases as the universe expands, it is more than surprising that the two are comparable just at the present time, while their ratio has to be infinitesimal in the early universe.

This review is organized as follows. In Section 2 we give an abreviated history of the $\Lambda$-term. The first aspect of the cosmological constant problem, related to fundamental physics, is stated more precisely in Section 3. In the next Section we turn to the recent evidence for an accelerating universe, coming from data on the Hubble diagram for type Ia supernovae. The cosmological significance of the temperature fluctuations of the cosmic microwave background radiation is explained in Section 5. We shall see that existing data suggest a spatially flat universe, but the uncertainties are still considerable. We conclude with a summary and some additional remarks.
\section{On the history of the $\Lambda$-term}
The cosmological term was introduced by Einstein when he applied general relativity the first time to cosmology. In his work \cite{Ein17} of 1917 he found the first cosmological solution of a consistent theory of gravity. This bold step can be regarded as the beginning of modern cosmology. 

In his attempt Einstein assumed that space is globally {\it closed}, because he then believed that this was the only way to satisfy Mach's principle, i.e., that the metric field should be determined uniquely by the energy-momentum tensor. In addition, Einstein assumed that the universe was {\it static}. This was very reasonable, because the relative velocities of the stars as observed at the time were extraordinarily small in comparison to the velocity of light. (Remember, astronomers only learned later that spiral nebulae are independent star systems outside the Milky Way. This was definitely established when Hubble found in 1924 that there were Cepheid variables in Andromeda and also in other galaxies. Five years later he announced the recession of galaxies.)

These two assumptions were, however, not compatible with Einstein's original field equations. For this reason, Einstein added the famous $\Lambda$-term which is compatible with general invariance and the energy-momentum law $\nabla_\nu T^{\mu \nu}=0$ for matter. The modified field equations are in standard notation \cite{Str85}:
\begin{equation}
G_{\mu \nu} \, = \, 8\pi G T_{\mu \nu} + \Lambda g_{\mu \nu}.
\label{eq:einstein}
\end{equation}
For the static Einstein universe these equations imply the two relations
\begin{equation}
8\pi G \rho \, = \, \frac{1}{a^2} \, = \, \Lambda,
\end{equation}
where $\rho$ is the mass density of the dust filled universe (zero pressure) and $a$ is the radius of curvature. (The geometry of space is necessarily a 3-sphere with radius $a$.) Einstein was very pleased by this direct connection between the mass density and geometry, because he thought that this was in accord with Mach's philosophy.

In the same year, 1917, de Sitter discovered a completely different static cosmological solution which also incorporated the cosmological constant, but was ``anti-Machian'', because it contained absolutely no matter. Until about 1930 almost everybody ``knew'' that the universe was static, in spite of the two important papers by Friedman in 1922 and 1924 and Lemaitre's work in 1927. These path-breaking papers were in fact largely ignored. In comments to Lemaitre, Einstein rejected the expanding universe solution as late as in 1928. It is also not well-known that Hubble interpreted his famous results on the redshift of the radiation emitted by distant nebulae in the framework of the de Sitter model. However, Lemaitre's successful explanation of Hubble's discovery finally changed the viewpoint of the majority of workers in the field. At this point Einstein rejected the cosmological term as superfluous and no longer justified \cite{Ein31}. Later he called its introduction as ``the biggest blunder of my life''. In this he was probably wrong.

After the ``$\Lambda$-force'' was rejected by its inventor, other cosmologists, like Eddington, retained it. One major reason was that it solved the problem of the age of the universe when the Hubble period was thought to be only 2 billion years (corresponding to the value $H_o \simeq 500$ km s$^{-1}$ Mpc$^{-1}$ of the Hubble constant). This was even shorter than the age of the Earth. In addition, Eddington and others overestimated the age of stars and stellar systems.

For this reason, the $\Lambda$-term was employed again and a model was revived which Lemaitre had singled out from the many possible solutions of the Friedmann-Lemaitre equations. This so-called Lemaitre hesitation universe is closed and has a repulsive $\Lambda$-force ($\Lambda >0$), which is slightly greater than the value chosen by Einstein. It begins with a big bang and has the following two stages of expansion. In the first the $\Lambda$-force is not important, the expansion is decelerated due to gravity and approaches slowly the radius of the Einstein universe. At about this time, the repulsion becomes stronger than gravity and a second stage of expansion begins which eventually inflates into a whimper. In this way a positive cosmological constant was employed to reconcile the expansion of the universe with the ages of stars.

The repulsive effect of a positive cosmological constant can be seen from the following consequence of Einstein's field equations for the time dependent scale factor $a(t)$:
\begin{equation}
\ddot{a} \, = \, - \frac{4 \pi G}{3} (\rho + 3p) a + \frac{\Lambda}{3} a,
\end{equation}
where $p$ is the pressure of all forms of matter.
\section{The cosmological constant problem}
Observationally the cosmological constant is known to be tiny in any natural scale of units. An upper bound is obtained as follows:

From Eq.(\ref{eq:einstein}) we see that the $\Lambda$-term gives an effective vacuum energy density contribution of magnitude
\begin{equation}
\rho_{\Lambda} \, = \, \Lambda / 8\pi G.
\end{equation}
From all we know, this density cannot be much larger than the critical density
\begin{eqnarray}
\rho_{crit} \, & = & \, \frac{3 H_o^2}{8 \pi G} \nonumber \\
& = & \, 1.88 \times 10^{-29} \, h_o^2 \, ({\rm g/cm}^3) \nonumber \\
& = & \, 8 \times 10^{-47} \, h_o^2 \, ({\rm GeV})^4,
\label{eq:obsbound}
\end{eqnarray}
where the ``reduced Hubble constant'' $h_o \, \equiv \, H_o /(100$ km s$^{-1}$ Mpc$^{-1}$) is in the range 0.65 $\pm$ 0.1. (Following general practice, we do not write the conversion factors involving $c$ and $\hbar$.) Thus 
\begin{equation}
\rho_{\Lambda} \, \leq \, 10^{-46} \, ({\rm GeV})^4
\end{equation}
and
\begin{equation}
\Lambda \, \leq \, 8 \pi G \rho_{crit} \, \simeq \, 10^{-120} \, M_{pl}^2,
\end{equation}
where $M_{pl}$ denotes the Planck mass. The smallness of this upper bound is a complete mystery for the following reasons.

Invariance arguments imply that the vacuum expectation value of the energy-momentum tensor has the form
\begin{equation}
< T_{\mu \nu} > \, = \, <\rho >_{vac} \, g_{\mu \nu}
\end{equation}
(ignoring higher curvature terms) and thus has the same effect as the cosmological term in (\ref{eq:einstein}). This implies that the {\it effective} cosmological constant, which controls the large scale behavior of the universe, is given by
\begin{equation}
\Lambda \, = \, 8 \pi G < \rho >_{vac} + \Lambda_0,
\label{eq:lambda}
\end{equation}
where $\Lambda_0$ is a bare cosmological constant. It is completely mysterious why the two terms on the right hand side should almost exactly cancel. This is, more precisely, the $\Lambda$-problem.

We expect that the early universe passed through a series of phase transitions, which are associated with various symmetry breakings. The structure of the vacuum thereby changed, and the corresponding changes of $< \rho >_{vac}$ should be gravitationnally relevant. For instance, the change of the energy density in the QCD phase transition is about $\Lambda^4_{QCD}/16 \pi^2 \simeq (10^{-1}$ 
GeV$)^4$, and this is {\it more than 40 orders of magnitudes larger} than the observational bound (\ref{eq:obsbound}). For the breaking of the electroweak gauge symmetry and other transitions at higher energy, the discrepancy is even much worse. The real question is: {\it Why does the vacuum energy of our present asymmetric vacuum state almost exactly vanish (more precisely, compensate a possible bare cosmological constant) ?}

This question was basically already asked by W. Pauli very early in his professional career, as I learned from some of his later assistants. Pauli wondered whether the zero-point energy of the electromagnetic field could be gravitationally effective. In those days the classical electron radius was considered to be a natural cut-off, and thus Pauli got for the vacuum energy density (in units with $\hbar=c=1$)
\begin{eqnarray}
<\rho >_{vac} \, & = & \, \frac{8 \pi}{(2\pi)^3} \, \int_0^{\omega_{max}} \, \frac {\omega}{2}~ \omega^2 \, d\omega \nonumber \\
& = & \, \frac{1}{8 \pi^2} \, \omega_{max}^4,
\end{eqnarray}
with 
\begin{equation}
\omega_{max} \, = \, \frac{2 \pi}{\lambda_{max}} \, = \, \frac{2 \pi m_e}{\alpha}.
\end{equation}
The corresponding radius of the Einstein universe in (2) would then be
\begin{equation}
a \, = \, \frac{\alpha^2}{(2 \pi)^{2/3}} \, \frac{M_{pl}}{m_e} \, \frac{1}{m_e} \, \simeq \, 31 \, {\rm km}.
\end{equation}
Pauli was quite amused to find that this universe ``would not even reach out to the moon''. Nowadays we use for $\omega_{max}$ at least 100 GeV, since we know that quantum fluctuations do not cancel in the standard model of particle physics, because there is no symmetry requiring this.

Another interesting remark was made by Zeldovich in 1967 during the third renaissance period of the $\Lambda$-term. First, he assumed - completely ad hoc - that the zero-point contributions to $<\rho>_{vac}$ are exactly cancelled by the bare term in (\ref{eq:lambda}). There remain then still higher-order effects. In particular, gravitational interactions between particles in the vacuum fluctuate. Dimensionally one would expect that these lead to a gravitational self-energy density of order $G\mu^6$, where $\mu$ is some cut-off mass \footnote{This is the gravitational self-energy of $(1/\mu)^{-3}$ particles with energy $\mu$ per unit volume.}. Even for $\mu$ as low as 1 GeV (for no good reason) this is about 9 orders of magnitude larger than the observational bound (\ref{eq:obsbound}). Hence even extreme fine tuning is destroyed by higher orders. This illustrates once more that there must be a deep reason for the smallness of the cosmological constant. The problem can also be phrased as follows: It is hard to see how general invariance can be broken such that the Lorentz group survives as a symmetry group.
\section{ Luminosity-redshift relation for type Ia supernovae}
The Hubble diagram is a graphic representation of the luminosity distance (magnitude) to some class of objects (e.g. type Ia supernovae) as a function of their redshift. Before presenting and discussing the exciting recent results we recall some basic facts.
\subsection{Theoretical background}
In cosmology several different distance measures are in use. They are all related by simple redshift factors. The one which is relevant in this Section is the {\it luminosity distance} $D_L$, defined by
\begin{equation}
D_L \, = \, ({\cal L}/4 \pi {\cal F})^{1/2} ,
\end{equation}
where ${\cal L}$ is the intrinsic luminosity of the source and ${\cal F}$ is the observed flux. Like all cosmological distances, $D_L$ is proportional to $c/H_o= 3000$ h$_o^{-1}$ Mpc. For Friedmann-Lemaitre models the ``Hubble-constant-free'' luminosity distance ${\cal D}_L$, defined by
\begin{equation}
D_L(z) \, = \, \frac{c}{H_o} \, {\cal D}_L(z;\Omega_M; \Omega_\Lambda) , 
\label{eq:lumdist}
\end{equation}
is a known dimensionless function of the redshift $z$, which depends parametrically on 
\begin{equation}
\Omega_M \, = \, \rho_M / \rho_{crit}~, \ \ \Omega_\Lambda =\rho_{\Lambda} / \rho_{crit}~,
\end{equation}
where $\rho_M$ is the total non-relativistic matter density (the relativistic matter of the present universe is negligible). The quantity $\Omega_K := 1-
\Omega_M -\Omega_\Lambda$ is a measure of the curvature. From the Friedmann equation
\begin{equation}
\left( \frac{\dot{a}}{a} \right) + \frac{k}{a^2} \, = \, \frac{8 \pi G}{3} \, 
(\rho + \rho_{\Lambda}) ,
\end{equation}
$k$ being the curvature constant, one finds $\Omega_K = -k / (a_o H_o)^2$. In particular, for $\Omega_K=0$ ($\Omega_M + \Omega_\Lambda =1)$ space is flat.

Astronomers use as logarithmic measures of ${\cal L}$ and ${\cal F}$ the absolute and apparent magnitudes \footnote{Beside the (bolometric) magnitudes $m,M$, astronomers use also magnitudes $m_B, m_V$,... refering to certain wave length bands $B$ (blue), $V$ (visual), and so on.}, denoted by $M$ and $m$, respectively. The conventions are chosen such that the distance modulus $m-M$ is related to $D_L$ as follows
\begin{equation}
m-M \, = \, 5~ {\rm log}\, \left( \frac{D_L}{1 {\rm Mpc}}\right) + 25~ .
\end{equation}
Inserting the representation (\ref{eq:lumdist}), we obtain the following relation between the apparent magnitude $m$ and the redshift $z$:
\begin{equation}
m \, = \, {\cal M} + 5~ {\rm log} \, {\cal D}_L (z;\Omega_M; \Omega_\Lambda),
\label{eq:m}
\end{equation}
where, for our purpose, ${\cal M} = M-5{\rm log} \, H_o -25$ is an uninteresting fit parameter. The comparison of this theoretical expectation with data will lead to interesting restrictions for the cosmological parameters $\Omega_M$ and $\Omega_\Lambda$.

In this context the following remark is important. For a fixed $z$ in the interesting interval [0.5,1.0], say, the equations ${\cal D}_L(z;\Omega_M, \Omega_\lambda)={\rm const}$ define degeneracy curves in the $\Omega$-plane. Since the curvature of these contours turns out to be small, we can associate an approximate slope to them. For $z=0.4$ the slope is about 1 and increases to $1.5-2$ by $z=0.8$ over the range of $\Omega_M$ and $\Omega_{\Lambda}$ of interest. Thus even very accurate data can at best select a narrow strip in the $\Omega$-plane, which is parallel to the degeneracy curves. This is the reason for the form of the likelyhood regions shown in Fig.2.
\subsection{Type Ia supernovae as standard candles}
It has long been recognized that supernovae of type Ia are excellent standard candles and are visible to cosmic distances of about 500 Mpc \cite{Baa38}. At relatively close distances they can be used to measure the Hubble constant, by calibrating the absolute magnitude of nearby supernovae with various distance determinations (e.g., Cepheids). There is still some dispute over these calibrations, resulting in differences of about 10\% for $H_0$. (For a review, see \cite{Tam98}).

In 1979 Tammann \cite{Tam79} and Colgate \cite{Col79} independently suggestet that at higher redshifts this class of supernovae can be used to determine also the deceleration parameter. In recent years this program became feasible thanks to the developement of new technologies which made it possible to obtain digital images of faint objects over sizable angular scales and by making use of big telescopes such as Hubble and Keck.

There are two major teams investigating high-redshift SNe Ia, namely the ``Supernova Cosmology Project'' (SCP) and the ``High-$Z$ Supernova Search Team'' (HZT). Each team has already found over 70 SNe, and both groups have published almost identical results.

Before discussing these, a few remarks about the nature and properties of type Ia SNe should be made. The immediate progenitors are most probably carbon-oxygen white dwarfs in close binary systems. In the standard scenario a white dwarf accretes matter from a nondegenerate companion until it approaches the critical Chandrasekhar mass and ignites carbon deep in its interior. This is followed by an outward-propagating subsonic nuclear flame (a deflagration) leading to a total disruption of the white dwarf. Within a few seconds the star is converted largely into nickel as well as other elements between silicon and iron. The dispersed nickel radioactively decays to cobalt and then to iron in a few hundred days. Since the physics of thermonuclear runaway burning in degenerate matter is complex and not really well understood, numerical simulations are not yet reliable in all details.

In some cases it may well be that a type Ia supernova is the result of a merging of two carbon-rich white dwarfs. However, no double-dwarf system with the necessary mass and orbital period has ever been identified. Moreover, simulations indicate that white dwarf mergers might not lead to strong nuclear explosions.

SNe Ia are not perfect standard candles. Their peak absolute magnitudes have a dispersion of 0.3-0.5 mag, depending on the sample. Astronomers have, however, learned in recent years to reduce this dispersion by making use of the light-curve shape. Examination of nearby SNe showed that the peak brightness is correlated with the time scale of their brightening and fading: Slow decliners tend to be brighter than the rapid ones. There are also some correlations with spectral properties. By making use of these correlations it became possible to reduce the remaining intrinsic dispersion to $\simeq 0.17$ mag. Other corrections, such as Galactic extinction, have been applied, resulting for each supernova in an effective (rest-frame) $B$ magnitude $m_B^{eff}$. The redshift dependence of this quantity is compared with the theoretical expectation (\ref{eq:m}).
\subsection{Results}
In Fig.1 the Hubble diagram $m_B^{eff}(z)$ for 42 high-redshift supernovae, published by the SCP team is shown, along with 18 Cal\'an/Tololo low-redshift supernovae \cite{Per99}. (For the HZT results, see \cite{Rie99}.)

The main result of the analysis is presented in Fig.2. The confidence regions imply in particular that $\Omega_{\Lambda}$ is {\it nonzero} at the $3\sigma$ statistical confidence level. An approximate fit is
\begin{equation}
   0.8\Omega_M-0.6\Omega_{\Lambda}\simeq -0.2\pm 0.1~.
\label{eq:fit}
\end{equation}
Assuming a flat model the data imply
\begin{equation}
   \Omega_M^{flat}=0.28_{-0.08}^{+0.09}~~(1\sigma~\rm{statistical})~~_{-0.04}^{+0.05}~~(\rm{identified~systematics})~.
\end{equation}
\subsection{Systematic uncertainties}
The results of both groups are limited by systematic errors which are seriously addressed. Of those that have been quantified so far, none appears to reconcile the data with $\Omega_{\Lambda}=0$. However, much more work is still necessary to make sure that there are indeed no systematic effects that might invalidate the results shown in Fig.2.

Perhaps the most serious uncertainty is {\it intrinsic evolution} of SNe Ia with redshift. Evolutionary effects are indicated by observations of nearby SNe: There is a mean luminosity difference in late-type and early-type host galaxies of about $\sim 0.3$ mag. On the basis of numerical simulations such differences are expected. They are corrected for by making use of the correlations mentioned earlier. So far there is no sign that the local calibration suffers a systematic drift with distance, but the last word has not been said.

{\it Extinction} is also an important concern. For a discussion of this and other systematic uncertainties we refer the reader to the original papers \cite{Per99},\cite{Rie99}.
\section{Cosmic microwave background anisotropies}
The observed properties (spectrum, anisotropies) of the cosmic microwave background (CMB) radiation give us direct information how the universe looked at the time of recombination (redshift $\simeq$ 1100). Indeed, on its way from the 'cosmic photosphere' the radiation suffered only small disturbances (e.g., by gravitational lensing).

From the observed CMB temperature anisotropies over the sky we can infer the spectrum of density fluctuations. These tiny perturbations evolved afterwards by gravitational amplification to the clumpy large scale we see today.

It is straightforward, although somewhat complicated, to analyze the evolution of temperature and density fluctuations {\it before} recombination. Since these are so tiny we are dealing with {\it linear} physics. We are allowed to linearize all basic equations (Einstein's field equation, Boltzmann's equation for the photons, and the fluid equations for the various components of matter) around the Friedmann-Lemaitre behavior. This leads to a system of ordinary differential equations that governs the time evolution of the amplitudes in a mode decomposition. However, what we do not know a priori are the initial conditions at some time long before recombination. We can, of course, work out the results of various possibilities and confront them with observations.

On the other hand, inflationary models of the very early universe strongly favor a cold dark matter (CDM) scenario with {\it adiabatic} initial fluctuations, whose spectrum is nearly {\it scale invariant}. It is remarkable that these properties are the result of unavoidable quantum fluctuations in the very early universe. For a CDM scenario with primordial adiabatic fluctuations a sequence of (acustic) peaks for the CMB angular power spectrum is predicted. Remarkably the angular position of the first peak depends mainly on the curvature parameter $\Omega_K$. (This position corresponds roughly to the ``sound horizon'', i.e., the distance that pressure waves can travel until recombination.) In other words, once the observations definitely reveal a first acustic peak, its angular position tells us the value of $\Omega_M+\Omega_{\Lambda}$. This information is largely complementary to (\ref{eq:fit}).

At the moment the observational situation is very fluent. There is increasing evidence for a first peak at about $1^{\circ}$ separation on the sky, which corresponds roughly to a flat universe ($\Omega_M+\Omega_{\Lambda}=1)$. Within a year or so, we will know more. (The current situation is discussed, for instance, in \cite{Bah99}.) Very accurate maps will be produced in the coming years by the NASA Microwave Anisotropy Probe (MAP) and the ESA PLANCK satellite missions.
\section{Concluding remarks}
During recent years it has become increasingly convincing that the average density of nonrelativistic matter in the universe is dominated by exotic forms, perhaps weakly interacting massive particles. The main evidence results from rich cluster data, in particular from the analysis of the X-radiation emitted by the hot intracluster gas. The total matter distribution of a rich cluster can also be determined by making use of gravitational lensing. Both methods support each other quite nicely.

The average matter density - exotic forms included - is, however, undercritical. Several independent methods lead to the value
$$
   \Omega_M\simeq 0.3\pm 0.1~.
$$
One arrives at this, for instance, if the information just mentioned is supplemented by the primordial abundances of the light elements, synthesized in the hot big bang. (For a discussion of other methods, we refer to \cite{Bah99}, and references therein.)

In Section 4 and 5 we have presented the evidence for a cosmologically significant vacuum energy density and a spatially flat universe. It now appears that we are living in a critical universe in which the vacuum energy dominates, and ordinary matter is only a tiny fraction:
$$
   \Omega_{\Lambda}\simeq 2/3,~\Omega_M\simeq 1/3,~\Omega_M>>\Omega_B
$$
($\Omega_B$ denotes the baryon fraction).
It should be stressed once more that this cannot be considered as definitely established. However, the present evidence is quite strong. The near future should be exciting.

Possible ways to avoid the cosmic coincidence problem have been discussed a lot recently. The general idea is to explain the accelerating expansion of the universe by yet another form of exotic missing energy with negative pressure, called {\it quintessence}. In concrete models this is described by a scalar field, whose dynamics is such that its energy naturally adjust itself to be comparable to the matter density today for generic initial conditions \cite{Ste98}.

\newpage
\begin{figure}
\leavevmode
    \epsfig{file=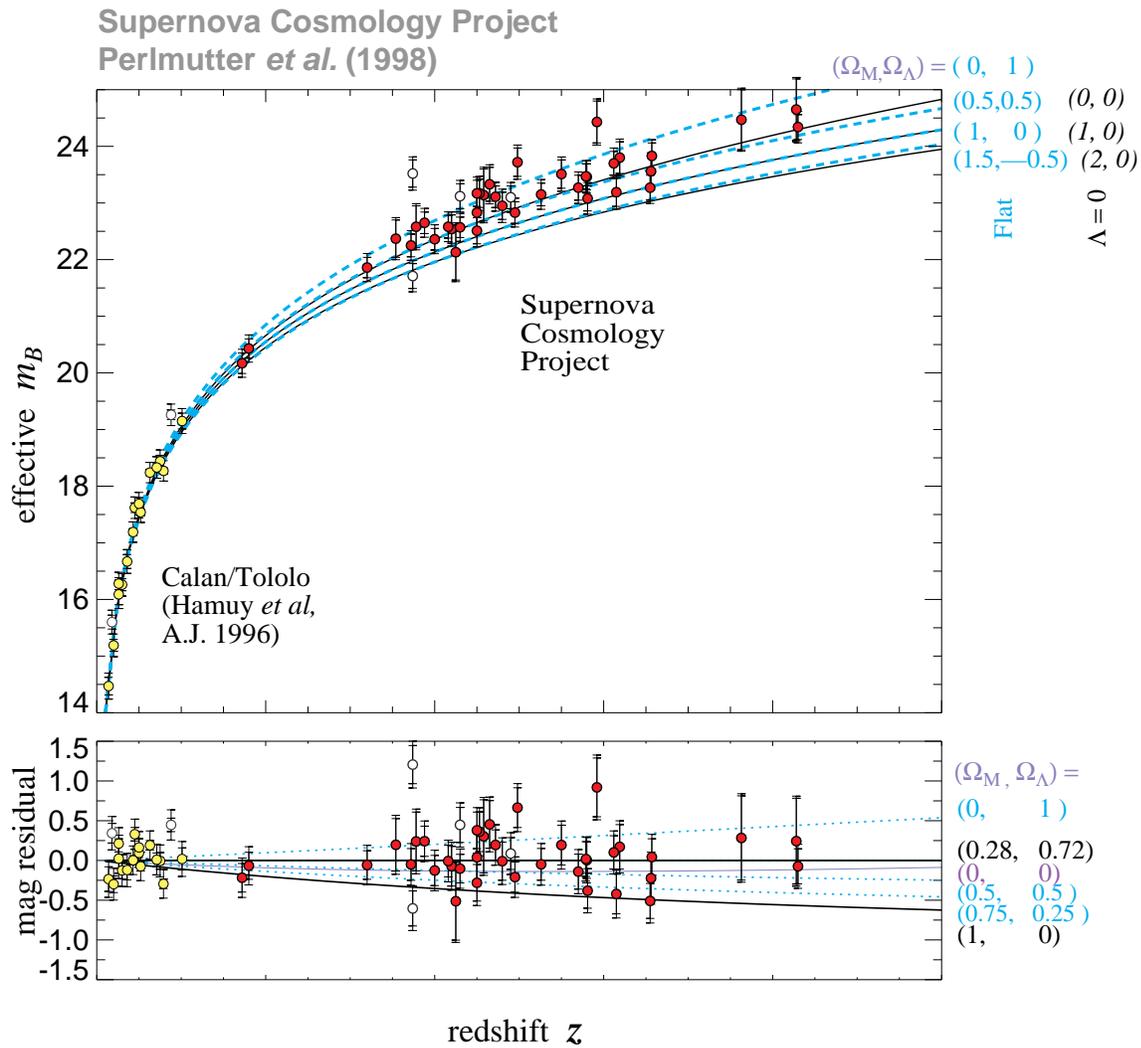,width=15cm,angle=0}
    \caption{(a) Magnitude-redshift relation for 42 high-redshift Type Ia supernovae from SCP, and 18 low-redshift supernovae from the Cal\'an/Tololo Survey. The solide curves correspond to the model expectations (\ref{eq:fit}) for $\Omega_{\Lambda}=0$, while the dashed curves are for a range of flat models. (b) The magnitude residuals from the best-fit flat model $(\Omega_M,\Omega_{\Lambda})=(0.28,0.72)$. (Adapted from Ref. \cite{Per99}.)}
\end{figure}

\newpage

\begin{figure}
\leavevmode
    \epsfig{file=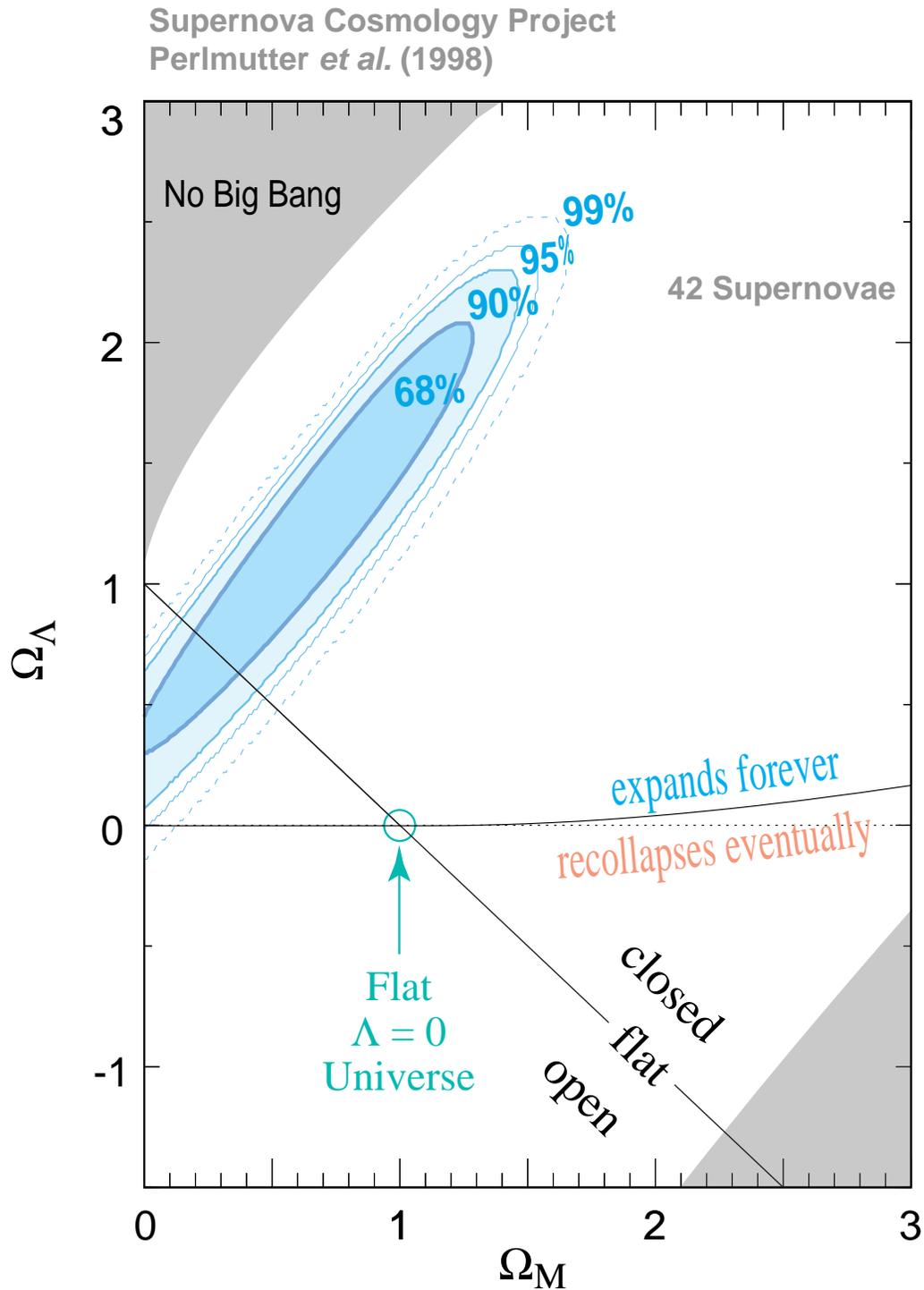,width=14cm,angle=0}
 \caption{Best-fit confidence regions in the $\Omega_M-\Omega_{\Lambda}$ plane. The upper-left shaded region represents ``bouncing models'' with no big bang in the past. The lower-right shaded region is excluded because the age of the universe would be too low. (From Ref. \cite{Per99}.)}
\end{figure}

\end{document}